\documentclass[12pt]{article}
\usepackage{amsmath}
\usepackage{graphicx}
\usepackage{color}
\begin{document}
\baselineskip=18 pt
\begin{center}
{\large{\bf Klein-Gordon oscillator in the presence of external fields in a cosmic space-time with a spacelike dislocation and Aharonov-Bohm effect}}
\end{center}

\vspace{.5cm}

\begin{center}
{\bf Faizuddin Ahmed}\footnote{faizuddinahmed15@gmail.com ; faiz4U.enter@rediffmail.com}\\ 
{\bf Ajmal College of Arts and Science, Dhubri-783324, Assam, India}
\end{center}

\vspace{.5cm}

\begin{abstract}

In this paper, we study interactions of a scalar particle with electromagnetic potential in the background space-time generated by a cosmic string with a spacelike dislocation. We solve the Klein-Gordon oscillator in the presence of external fields including an internal magnetic flux field and analyze the analogue effect to the Aharonov-Bohm effect for bound states. We extend this analysis subject to a Cornell-type scalar potential, and observe the effects on the relativistic energy eigenvalue and eigenfunction.

\end{abstract}

{\bf keywords:} Relativistic wave-equation : bound states, electromagnetic interactions, Aharonov-Bohm effect, energy spectrum, wave-function, topological defects, scalar potential.

\vspace{0.2cm}


\section{Introduction}

The Klein-Gordon oscillator \cite{aa1,aa2} was inspired by the Dirac oscillator \cite{aa3} applied to spin-$\frac{1}{2}$ particles. The Klein-Gordon oscillator has been investigated in several physical systems, such as in the background of the cosmic string with external fields \cite{aa4}, in the presence of Coulomb-type potential considering two ways: (i) by modifying the mass term $m \rightarrow m+S$ \cite{aa5}, (ii) via the minimal coupling \cite{aa6} with a linear potential, in the background space-time produced by topological defects using the Kaluza-Klein theory \cite{aa7}, in the Som-Raychaudhuri space-time in the presence of external fields \cite{aa8}, motion of electron in an external magnetic field in the presence of screw dislocations \cite{aa9}, continuous distribution of screw dislocation \cite{aa10}, in the presence of Cornell-type potential in a cosmic string space-time \cite{aa11}, relativistic quantum dynamics of DKP oscillator field subject to a linear scalar potential \cite{aa12}, DKP equation for spin-zero bosons subject to a linear scalar potential \cite{aa13}, and the Dirac equation subject to a vector and scalar potentials \cite{aa14}. In literature, it is known that cosmic string has been produced by phase transitions in the early universe \cite{aa15} as it is predicted in the extensions of the standard model \cite{aa16,aa17}. Topological defects in condensed matter physics can be associated with the presence of curvature and torsion. In particular, topological effects associated with torsion have been investigated in crystalline solids with the use of differential geometry \cite{aa18,aa19}. Recent studies have explored the effects of torsion on condensed matter systems \cite{aa20,aa21,aa22,aa23}. Therefore, there is a great interest in the connection between quantum mechanics and the general relativity.

The cosmic string space-time in cylindrical coordinates $(t, r, \phi, z)$ is described by the following line element \cite{aa16,aa17,aa18,aa24,aa25,aa26,aa27,aa28,aa29}
\begin{equation}
ds^2=-dt^2 + dr^2 +\alpha^2\,r^2\,d\phi^2+dz^2,
\label{cosmic}
\end{equation}
where $\alpha=(1-4\,\mu)$ is the topological parameter with $\mu$ being the linear mass density of cosmic string, and $ 0 < \alpha < 1$. Furthermore, in the cylindrical symmetry we have that $ 0 < r \leq \infty$, $ 0 \leq \phi \leq  2\pi$ and $-\infty < z < \infty$.

In Ref. \cite{aa30}, Klein-Gordon oscillator without and/or with a linear scalar potential in the presence of external fields including an internal magnetic flux field in a space-time with a spacelike dislocation were studied. They solved the wave-equation analytically and analyzed the effects on the relativistic energy eigenvalue. In Ref. \cite{aa31}, authors investigated the Klein-Gordon oscillator with topological defects including an internal magnetic flux field subject to a Coulomb-type plus linear potential (called Cornell-type potential) in a space-time with screw dislocation (space-like dislocation). They obtained the relativistic energy eigenvalue and observed the analogous effect to the Aharonov-Bohm effect for bound states. In this work, we study the Klein-Gordon oscillator field interacts with external fields including an internal magnetic flux field in a space-time with a magnetic screw dislocation. We extend this analysis subject to a Cornell-type scalar potential by searching analytical for bound states solution to the system. Recently, dislocation has been investigated in nonrelativistic and relativistic quantum systems. The spiral dislocation, in the nonrelativistic context, has been investigated in the harmonic oscillator \cite{aa33}; in the relativistic context, it has been investigated in a scalar field in a noninertial frame \cite{aa34}. The screw dislocation, in the nonrelativistic context, has been applied in the harmonic oscillator \cite{aa35,aa36}, in the Landau quantization \cite{aa9,aa10,aa37}, in the doubly anharmonic oscillator \cite{aa38}, in Landau quantization for an induced electric dipole \cite{aa39}, and in noninertial effects on a nonrelativistic Dirac particle \cite{aa40}. In relativistic context, the screw dislocation has been studied in the Dirac oscillator \cite{aa41,aa42}, in the Klein-Gordon oscillator \cite{aa31}, and analogue effects to the Aharonov-Bohm effect for bound states in a position-dependent mass system \cite{RLLV2}.

The structure of the present paper is as follow: in {\it section 2.1}, we investigate the Klein-Gordon oscillator in the presence external field including an internal magnetic flux a cosmic string space-time with a spacelike dislocation; in {\it section 2.2}, we extend this analysis subject to a Cornell-type scalar potential and,  analyze the analogue effect to the Aharonov–Bohm effect for bound states, and finally conclusions in {\it section 3}.

\section{Klein-Gordon oscillator interacts with external fields in a cosmic space-time with a spacelike dislocation}

Let us begin this section by introducing the space-time with a screw dislocation. It corresponds to a space-time with a linear topological defect associated with torsion and cosmic string that can be described by the line element \cite{aa30}
\begin{equation}
ds^2=-dt^2+dr^2+\alpha^2\,r^2\,d\phi^2+(dz+\chi\,d\phi)^2,
\label{1}
\end{equation}
where $c=1=\hbar$, $ 0 < \alpha < 1$ is the cosmic string parameter, and $\chi$ is the dislocation (torsion) parameter and in condensed matter physics, this parameter is related to the Burgers vector {\bf b}  via $\chi=\frac{b}{2\,\pi}$ \cite{aa18,aa19,aa44,aa45}. It is important to mention that the screw dislocation (torsion) corresponds to a singularity at the origin \cite{aa18,aa44}. Also note that the spatial part of the metric (\ref{1}) is called the Katanaev and Volovich line element in studies of solids as the screw dislocation \cite{aa18,aa19}. For $\chi \rightarrow 0$, the metric (\ref{1}) reduces to a cosmic string space-time. Again for $\chi \rightarrow 0$, and $\alpha \rightarrow 1$, the space-time reduces to Minkowski flat space metric in cylindrical coordinates system.

The metric tensor for the space-time (\ref{1}) to be 
\begin{equation}
g_{\mu\nu} ({\bf x})=\left (\begin{array}{llll}
-1 & 0 & \quad 0 & 0 \\
0 & 1 & \quad 0 & 0 \\
0 & 0 & \alpha^2\,r^2+\chi^2 & \chi \\
0 & 0 & \quad \chi & 1
\end{array} \right)
\label{2}
\end{equation}
with its inverse 
\begin{equation}
g^{\mu\nu} ({\bf x})=\left (\begin{array}{llll}
-1 & 0 & \quad 0 & \quad 0 \\
\quad 0 & 1 & 0 & \quad 0 \\
\quad 0 & 0 & \frac{1}{\alpha^2\,r^2} & -\frac{\chi}{\alpha^2\,r^2} \\
\quad 0 & 0 & -\frac{\chi}{\alpha^2\,r^2} & 1+\frac{\chi^2}{\alpha^2\,r^2}
\end{array} \right)
\label{3}
\end{equation}
The metric has signature $(-,+,+,+)$ and the determinant of the corresponding metric tensor $g_{\mu\nu}$ is
\begin{equation}
det\;g=-\alpha^2\,r^2.
\label{4}
\end{equation}

The relativistic quantum dynamics of charged particles of mass $m$ is described by the Klein-Gordon equation \cite{aa28}
\begin{equation}
[\frac{1}{\sqrt{-g}}\,D_{\mu} (\sqrt{-g}\,g^{\mu\nu}\,D_{\nu})-m^2]\,\Psi=0,
\label{5}
\end{equation}
where the minimal coupling with electromagnetic interaction is as follows
\begin{equation}
D_{\mu}=\partial_{\mu}-i\,e\,A_{\mu},
\label{6}
\end{equation}
where $e$ is the electric charge and $A_{\mu}$ is the electromagnetic four-vector potential defined by
\begin{equation}
A_{\mu}=(0,\vec{A})\quad,\quad \vec{A}=(0,A_{\phi},0).
\label{7}
\end{equation}
The three-vector potential in the symmetric gauge is defined by
\begin{equation}
\vec{A}=\vec{A}_1+\vec{A}_2,
\label{gauge}
\end{equation}
where the angular component of three-vector potential for $\vec{A}_1$ is \cite{aa28}
\begin{equation}
A^{(1)}_{\phi}=-\frac{1}{2}\,\alpha\,B_0\,r^2.
\label{gauge2}
\end{equation}
And that for $\vec{A}_2$ is  \cite{aa30,aa31,RLLV,RLLV2,cc4}
\begin{equation}
A^{(2)}_{\phi}=\frac{\Phi_B}{2\,\pi}
\label{gauge3}
\end{equation}
so that the angular component of the electromagnetic four-vector potential is
\begin{equation}
A_{\phi}=-\frac{1}{2}\,\alpha\,B_0\,r^2+\frac{\Phi_B}{2\,\pi}.
\label{14}
\end{equation}
The magnetic field is along the $z$-direction given by
\begin{equation}
\vec{B}=\vec{\nabla} \times \vec{A}=-B_{0}\,\hat{k}.
\label{gauge4}
\end{equation}
Here $\Phi_B=const$ is an internal quantum magnetic flux through the core of the topological defect \cite{aa35} and $e\,A_{\phi}=\Phi$ where $\Phi=\frac{\Phi_B}{(2\,\pi/e)}$. 

If one introduces a scalar potential by modifying the mass term as $m \rightarrow m+ S(r)$ into the above equation, we have
\begin{equation}
[\frac{1}{\sqrt{-g}}\,D_{\mu} (\sqrt{-g}\,g^{\mu\nu}\,D_{\nu})-(m+S)^2]\,\Psi=0.
\label{8}
\end{equation}
Using the geometry (\ref{1}), Eq. (\ref{8}) becomes
\begin{equation}
[-\partial_{t}^2+\frac{1}{r}\,\partial_{r}\,(r\,\partial_{r})+\frac{1}{\alpha^2\,r^2}\,(\partial_{\phi}-i\,e\,A_{\phi}-\chi\,\partial_{z})^2+\partial_{z}^2-(m+S)^2]\,\Psi=0.
\label{9}
\end{equation}

To include the oscillator with Klein-Gordon field, we change the following momentum operator \cite{aa7}:
\begin{equation}
\vec{p} \rightarrow \vec{p}+i\,m\,\Omega\,\vec{r},
\label{10}
\end{equation}
where $\Omega$ is the oscillator frequency and $\vec{r}=r\,\hat{r}$ with $r$ being distance from the particle to the string. So we can write $p^2 \rightarrow (\vec{p}+i\,m\,\Omega\,\vec{r})(\vec{p}-i\,m\,\Omega\,\vec{r}) $.

Therefore, Eq. (\ref{9}) becomes
\begin{eqnarray}
&&[-\partial_{t}^2+\frac{1}{r}\,(\partial_{r}+m\,\Omega\,r)\,(r\,\partial_{r}-m\,\Omega\,r^2)+\frac{1}{\alpha^2\,r^2}\,(\partial_{\phi}-i\,e\,A_{\phi}-\chi\,\partial_{z})^2\nonumber\\
&&+\partial_{z}^2-(m+S)^2]\,\Psi=0\nonumber\\\Rightarrow 
&&[-\partial_{t}^2+\frac{1}{r}\,\partial_{r}\,(r\,\partial_{r})-2\,m\,\Omega-m^2\,\Omega^2\,r^2+\frac{1}{\alpha^2\,r^2}\,(\partial_{\phi}-i\,e\,A_{\phi}-\chi\,\partial_{z})^2\nonumber\\
&&+\partial_{z}^2-(m+S)^2]\,\Psi=0.
\label{11}
\end{eqnarray}
We choose the following ansatz for the function $\Psi$
\begin{equation}
\Psi (t,r,\phi,z)=e^{i\,(-E\,t+l\,\phi+k\,z)}\,\psi (r),
\label{12}
\end{equation}
where $E$ is the energy, $l=0,\pm\,1,\pm\,2....\in {\bf Z}$ are the eigenvalue of the $z$-component of the angular momentum operator, and $k$ is a constant.

Substituting the above ansatz (\ref{12}) into the Eq. (\ref{11}), we have
\begin{eqnarray}
&&\psi''(r)+\frac{1}{r}\,\psi' (r)+[E^2-2\,m\,\Omega-m^2\,\Omega^2\,r^2-\frac{1}{\alpha^2\,r^2}\,(l-e\,A_{\phi}-k\,\chi)^2-k^2\nonumber\\
&&-(m+S)^2]\,\psi=0.
\label{13}
\end{eqnarray}

\subsection {Interaction without a scalar potential $S=0$}

Here we investigate the gravitational effect produced by the topological defects (cosmic string) on the above relativistic quantum system without scalar potential in the presence of external fields including an internal magnetic flux field. We see that the relativistic energy eigenvalue modify by the topological defects and break their degeneracy.

Substituting the angular component of four-vector potential Eq. (\ref{14}) into the Eq. (\ref{13}), we obtain the following differential equation
\begin{equation}
\psi'' (r)+\frac{1}{r}\,\psi' (r)+[\lambda-m^2\,\omega^2\,r^2-\frac{j^2}{r^2}]\,\psi=0,
\label{b1}
\end{equation}
where
\begin{eqnarray}
&&\lambda=E^2-k^2-\frac{2\,m\,\omega_c}{\alpha}\,(\alpha\,l_{eff}-k\,\chi)-m^2-2\,m\,\Omega,\nonumber\\
&&j=\frac{|l-\Phi-k\,\chi|}{\alpha},\nonumber\\
&&\omega=\sqrt{\omega^2_{c}+\Omega^2},\nonumber\\
&&l_{eff}=\frac{1}{\alpha}\,(l-\Phi),\nonumber\\ 
&&\Phi=\frac{\Phi_B}{(2\,\pi/e)}\nonumber \\ \mbox{and}
&&\omega_c=\frac{e\,B_0}{2\,m}
\label{b2}
\end{eqnarray}
is the cyclotron frequency.

Transforming $s=m\,\omega\,r^2$ into the above equation, we obtain the following differential equation \cite{aa47}
\begin{equation}
\psi'' (s)+\frac{1}{s}\,\psi' (s)+\frac{1}{s^2}\,(-\xi_1\,s^2+\xi_2\,s-\xi_3)\,\psi (s)=0,
\label{b3}
\end{equation}
where
\begin{equation}
\xi_1=\frac{1}{4}\quad,\quad \xi_2=\frac{\lambda}{4\,m\,\omega}\quad,\quad \xi_3=\frac{j^2}{4}.
\label{b4}
\end{equation}
The energy eigenvalue is given by
\begin{eqnarray}
&&(2\,n+1)\,\sqrt{\xi_1}-\xi_2+2\,\sqrt{\xi_1\,\xi_3}=0\nonumber\\\Rightarrow 
&&\lambda=2\,m\,\sqrt{\omega^2_{c}+\Omega^2}\,(2\,n+1+j)\nonumber\\\Rightarrow
&&E^2_{n,l}=m^2+k^2+2\,m\,\Omega+2\,m\,\sqrt{\omega^2_{c}+\Omega^2}\,(2\,n+1+\frac{|l-k\,\chi-\Phi|}{\alpha})\nonumber\\
&&+2\,m\,\omega_c\,\frac{(l-k\,\chi-\Phi)}{\alpha},
\label{b6}
\end{eqnarray}
where $n=0,1,2,3,4,...$.

The eigenfunction is given by
\begin{equation}
\psi (s)=s^{\frac{|l-\Phi-k\,\chi|}{2\,\alpha}}\,e^{-\frac{s}{2}}\,L^{(\frac{|l-\Phi-k\,\chi|}{\alpha})}_{n} (s),
\label{b7}
\end{equation}
where $L^{(\beta)}_{n} (s)$ is the generalized Laguerre polynomials.

If we take $\alpha \rightarrow 1$, one will recover the results obtained in \cite{aa30} (see Eq. (31) in \cite{aa30}). As the cosmic string parameter $\alpha$ are in the values $0 <\alpha < 1$. We can see that the presence of cosmic string parameter $\alpha$ modifies the energy spectrum. For $\Phi_B \neq 0$ and $\chi \neq 0$, we can observe in Eq. (\ref{b6}) that there exists an effective angular momentum, $l \rightarrow l_{eff}=\frac{1}{\alpha}\,(l-\Phi-k\,\chi)$. Thus the relativistic energy eigenvalue depend on the Aharonov-Bohm geometric quantum phase \cite{cc1}. This dependence of the energy eigenvalue on the geometric quantum phase gives rise to the analogous effect to the Aharonov-Bohm effect for bound states \cite{cc2,aa48,aa49}. Besides, we have that $E_{n,{\bar l}} (\Phi_B+\Phi_0)=E_{n,{\bar l} \mp \tau} (\Phi_B)$ where, $\Phi_0=\pm\,\frac{2\,\pi\,\alpha}{e}\,\tau$ with $\tau=1,2,3,.......$ and ${\bar l}=\frac{l}{\alpha}$.

For the zero torsion parameter, $\chi=0$, Eq. (\ref{b6}) becomes
\begin{eqnarray}
&&E^2_{n,l}=m^2+k^2+2\,m\,\Omega+2\,m\,\sqrt{\omega^2_{c}+\Omega^2}\,(2\,n+1+\frac{|l-\Phi|}{\alpha})\nonumber\\
&&+2\,m\,\omega_c\,\frac{(l-\Phi)}{\alpha}.
\label{b9}
\end{eqnarray}
Equation (\ref{b9}) is the energy spectrum of Klein-Gordon oscillator in the presence of external uniform magnetic field including an internal magnetic flux in a cosmic string space-time. For $\Phi_B \neq 0$, we can observe in Eq. (\ref{b9}) that the angular quantum number is shifted, $l \rightarrow l'=\frac{1}{\alpha}\,(l-\frac{\phi_B}{(2\,\pi/e)})$ and thus the relativistic energy eigenvalue depends on the Aharonov-Bohm geometric phase \cite{cc1}. This dependence of the energy eigenvalue on the geometric quantum phase gives rise to the analogous effect to the Aharonov-Bohm effect for bound states \cite{cc2,aa48,aa49}. Besides, we have that $E_{n,{\bar l}} (\Phi_B+\Phi_0)=E_{n,{\bar l} \mp \tau} (\Phi_B)$ where, $\Phi_0=\pm\,\frac{2\,\pi\,\alpha}{e}\,\tau$ with $\tau=1,2,3,.......$ and ${\bar l}=\frac{l}{\alpha}$. By taking $\Phi_B=0$ in Eq. (\ref{b9}), we have that the relativistic energy levels that arise from the interaction of the Klein-Gordon oscillator with a uniform magnetic field in the Minkowski space-time with a cosmic string. On the other hand, by taking $\omega_c \rightarrow 0$ and $\Phi_B \rightarrow 0$ in Eq. (\ref{b9}), we recover the results obtained in Ref. \cite{aa7}. Thus for $\chi \neq 0$ and $\Phi_B \neq 0$ in the energy eigenvalue Eq. (\ref{b6}), we have that the presence of torsion in the space-time modifies the degeneracy of the relativistic energy levels. Besides, the presence of torsion in the space-time changes the pattern of oscillation of the energy levels.

\subsection{Interaction with a Cornell-type scalar potential} 

Here we investigate the above relativistic quantum system described by the Klein-Gordon oscillator subject to a Cornell-type scalar potential in the presence of external fields including an internal magnetic flux field. A scalar potential is included into the systems by modifying the mass $m \rightarrow m+S (r)$ which is called position dependent mass system in the relativistic quantum systems (see {\it e. g.}, \cite{aa5,aa6,aa8,aa28,aa30,aa31,RLLV2,aa46,aa50,aa53,aa54,aa55,cc3,cc4,cc5,cc6,cc8,cc9,cc10}).

The Cornell potential, which consists of a linear potential plus a Coulomb potential, is a particular case of the quark-antiquark interaction, which has one more harmonic type term \cite{aa50}. Recently, the Cornell potential has been studied in the ground state of three quarks \cite{aa51}. However, this type of potential is worked on spherical symmetry; in cylindrical symmetry, which in our case, this type of potential is known as Cornell-type potential \cite{aa8,aa31,aa53,aa54,aa55}.

The Cornell-type scalar potential is given by
\begin{equation}
S=\frac{\eta_c}{r}+\eta_{L}\,r,
\label{16}
\end{equation}
where $\eta_c,\eta_L$ are the positive arbitrary potential parameters. This potential has been used successfully in models describing binding states of heavy quarks \cite{aa56,aa57,aa58}. The Cornell potential contains a short-range part dominated by a Coulombic term of quark and gluon interaction $\sim \frac{a}{r}$, and the large distance quark confinement as a linear term $\sim b\,r$ \cite{aa59,aa60,bb1,bb2,bb3,bb4}. In some situations when the parameter $b$ is small, it provides a particular case of perturbed Coulomb problem in atomic physics \cite{bb5}. This potential has been used to study the strange, charmed, and beautiful  baryon masses in the framework of variational approach \cite{bb6}.

Substituting the vector potential Eq. (\ref{14}) and the scalar potential Eq. (\ref{16}) into the Eq. (\ref{13}), we obtain the following differetial equation:
\begin{equation}
\psi'' (r)+\frac{1}{r}\,\psi'(r)+[\tilde{\lambda}-\tilde{\omega}^2\,r^2-\frac{\tilde{j}^2}{r^2}-\frac{a}{r}-b\,r]\,\psi (r)=0,
\label{17}
\end{equation}
where
\begin{eqnarray}
&&\tilde{\lambda}=E^2-m^2-k^2-2\,\eta_c\,\eta_L-\frac{2\,m\,\omega_c}{\alpha}\,(l-\Phi-k\,\chi)-2\,m\,\Omega,\nonumber\\
&&\tilde{\omega}=\sqrt{m^2\,(\omega^2_{c}+\Omega^2)+\eta^2_{L}},\nonumber\\
&&\tilde{j}=\sqrt{\frac{(l-\Phi-k\,\chi)^2}{\alpha^2}+\eta^2_{c}},\nonumber\\
&&a=2\,m\,\eta_c,\nonumber\\
&&b=2\,m\,\eta_L.
\label{function}
\end{eqnarray}

Transforming $x=\sqrt{\tilde{\omega}}\,r$ into the above Eq. (\ref{17}), we obtain the following wave equation:
\begin{equation}
\psi ''(x)+\frac{1}{x}\,\psi' (x)+[\zeta-x^2-\frac{\tilde{j}^2}{x^2}-\frac{\eta}{x}-\theta\,x]\,\psi (x)=0,
\label{18}
\end{equation}
where we have defined
\begin{equation}
\zeta=\frac{\tilde{\lambda}}{\tilde{\omega}}\quad,\quad \eta=\frac{a}{\sqrt{\tilde{\omega}}}\quad,\quad \theta=\frac{b}{\tilde{\omega}^{\frac{3}{2}}}.
\label{19}
\end{equation}

Now, we use the appropriate boundary conditions to investigate the bound states solution in this problem. It is require that the wave-functions must be regular both at $x \rightarrow 0$ and $x \rightarrow \infty $. These conditions are necessary since the wave-functions must be well-behaved in these limit, and thus, bound states of energy for the system can be obtained. Suppose the possible solution to the Eq. (\ref{18}) is
\begin{equation}
\psi (x)=x^{\tilde{j}}\,e^{-\frac{1}{2}\,(\theta+x)\,x}\,H (x),
\label{20}
\end{equation}
where $H (x)$ is an unknown function. Substituting the solution (\ref{20}) into the Eq. (\ref{18}), we obtain
\begin{equation}
H'' (x)+[\frac{\gamma}{x}-\theta-2\,x]\,H' (x)+[-\frac{\beta}{x}+\Theta]\,H (x)=0,
\label{23}
\end{equation}
where
\begin{eqnarray}
&&\gamma=1+2\,\tilde{j},\nonumber\\
&&\Theta=\zeta+\frac{\theta^2}{4}-2\,(1+\tilde{j}),\nonumber\\
&&\beta=\eta+\frac{\theta}{2}\,(1+2\,\tilde{j}).
\label{24}
\end{eqnarray}
Equation (\ref{23}) is the biconfluent Heun's differential equation \cite{aa28,bb7,bb8} with $H(x)$ is the Heun's polynomials function. Many authors studied the analytical solutions to the relativistic wave-equations in terms of Heun functions ({\it e. g.},\cite{bb9,bb10,bb11,bb12,bb13,bb14}).

The above equation (\ref{23}) can be solved by the series solution method. Writing the solution as a power series expansion around the origin \cite{bb15}:
\begin{equation}
H (x)=\sum^{\infty}_{i=0}\,c_{i}\,x^{i}.
\label{25}
\end{equation}
Substituting the power series solution (\ref{25}) into the Eq. (\ref{23}), we get the following recurrernce relation for the coefficients:
\begin{equation}
c_{n+2}=\frac{1}{(n+2)(n+2+2\,\tilde{j})}\,[\{\beta+\theta\,(n+1)\}\,c_{n+1}-(\Theta-2\,n)\,c_{n}].
\label{26}
\end{equation}
And the various coefficients are
\begin{eqnarray}
&&c_1=(\frac{\eta}{(1+2\,\tilde{j})}+\frac{\theta}{2})\,c_0,\nonumber\\
&&c_2=\frac{1}{4\,(1+\tilde{j})}\,[(\beta+\theta)\,c_{1}-\Theta\,c_{0}].
\label{27}
\end{eqnarray}

As the function $H (x)$ has a power series expansion around the origin in Eq. (\ref{25}), then, the relativistic bound states solution can be achieved by imposing that the power series expansion becomes a polynomial of degree $n$. Through the recurrence relation (\ref{26}), we can see that the power series expansion $H (r)$ becomes a polynomial of degree $n$ by imposing the following two conditions \cite{aa28}
\begin{eqnarray}
&&\Theta=2\,n,\quad (n=1,2,3,4,......)\nonumber\\
&&c_{n+1}=0.
\label{28}
\end{eqnarray}
By analysing the condition $\Theta=2\,n$, we get the following equation of eigenvalue $E_{n,l}$:
\begin{eqnarray}
&&E^2_{n,l}=m^2+k^2+2\,\eta_c\,\eta_L+2\,m\,\Omega+\frac{2\,m\,\omega_c}{\alpha}\,(l-k\,\chi-\Phi)\nonumber\\
&&+2\,\tilde{\omega}\,(n+1+\sqrt{\frac{(l-k\,\chi-\Phi)^2}{\alpha^2}+\eta^2_{c}})-\frac{m^2\,\eta^2_{L}}{\tilde{\omega}^2},
\label{29}
\end{eqnarray}
where $\tilde{\omega}$ is given in Eq. (\ref{function}).

For $\Phi_B \neq 0$ and $\chi \neq 0$, we can observe in Eq. (\ref{29}) that there exists an effective angular momentum quantum number, $l_{eff}=\frac{1}{\alpha}\,(l-k\,\chi-\frac{\phi_B}{(2\,\pi/e)})$. Thus the relativistic energy eigenvalue depend on the Aharonov-Bohm geometric phase \cite{cc1}. This dependence on the geometric quantum phase gives rise to the analogous effect to the Aharonov-Bohm effect for bound states \cite{cc2,aa48,aa49}. Besides, we have that $E_{n,{\bar l}} (\Phi_B+\Phi_0)=E_{n,{\bar l} \mp \tau} (\Phi_B)$ where, $\Phi_0=\pm\,\frac{2\,\pi\,\alpha}{e}\,\tau$ with $\tau=1,2,3,.......$ and ${\bar l}=\frac{l}{\alpha}$.

The wave-function is given by
\begin{equation}
\psi_{n,l} (x)=x^{\sqrt{\frac{(l-\Phi-k\,\chi)^2}{\alpha^2}+\eta^2_{c}}}\,e^{-\frac{1}{2}\,(\theta+x)\,x}\,H (x).
\label{33}
\end{equation}

Now, we impose additional recurrence condition $c_{n+1}=0$ to find the individual energy levels and corresponding wave-functions one by one as done in \cite{bb16,bb17}. As example, for $n=1$, we have $c_2=0$ which implies from (\ref{27})
\begin{equation}
c_1=\frac{2}{(\beta+\theta)}\,c_0 \Rightarrow \frac{\eta}{1+2\,\tilde{j}}+\frac{\theta}{2}=\frac{2}{(\beta+\theta)}
\label{34}
\end{equation}
a constraint on the physical parameters from which one can find $\tilde{\omega}_{1,l}$.

Therefore, the ground state energy level for $n=1$ from (\ref{29}) is given by
\begin{eqnarray}
&&E^2_{1,l}=m^2+k^2+2\,\eta_c\,\eta_L+2\,m\,\Omega+\frac{2\,m\,\omega_c}{\alpha}\,(l-k\,\chi-\Phi)\nonumber\\
&&+2\,\tilde{\omega_{1,l}}\,(2+\sqrt{\frac{(l-k\,\chi-\Phi)^2}{\alpha^2}+\eta^2_{c}})-\frac{m^2\,\eta^2_{L}}{\tilde{\omega}^2_{1,l}}.
\label{35}
\end{eqnarray}
With $\chi=0$, we have that the ground state energies (\ref{35}) becomes that stem from
the interaction of the Klein-Gordon oscillator with a magnetic field and a Cornell-type scalar
potential in the Minkowski space-time with a cosmic string, which is also a periodic function of the Aharonov-Bohm geometric quantum phase. Thus, the topology of the space-time also changes the pattern of oscillation of the ground state energies.

The corresponding ground state eigenfunctions using (\ref{33}) is given by
\begin{equation}
\psi_{1,l}=x^{\sqrt{\frac{(l-\Phi-k\,\chi)^2}{\alpha^2}+\eta^2_{c}}}\,e^{-\frac{1}{2}\,(x+\frac{2\,m\,\eta_{1,L}}{\tilde{\omega}_{1,l}^{\frac{3}{2}}})\,x}\,[1+\frac{1}{\sqrt{\tilde{\omega}_{1,l}}}\,\{\frac{2\,m\,\eta_c}{(1+2\,j)}+\frac{m\,\eta_{1,L}}{\tilde{\omega}_{1,l}} \}\,x],
\label{37}
\end{equation}
where the potential parameter $\eta_L \rightarrow \eta_{1,L}$ is so adjusted that the first order polynomial solution to the bound states can be obtained. Similarly one can evaluate the energy levels and wave-functions for $n=2,3$ and so on.

Now, we discuss below few cases of the above obtained energy eigenvalues.

\vspace{0.5cm}
{\bf Case 1 :} Linear scalar potential $S=\eta_L\,r$.
\vspace{0.5cm}

We consider here $\eta_c \rightarrow 0$, that is, only a linear scalar potential into the relativistic quantum system. 

Therefore, the energy eigenvalue Eq. (\ref{29}) becomes
\begin{eqnarray}
&&E^2_{n,l}=m^2+k^2+2\,m\,\Omega+\frac{2\,m\,\omega_c}{\alpha}\,(l-k\,\chi-\Phi)\nonumber\\
&&+2\,\tilde{\omega}\,(n+1+\frac{|l-k\,\chi-\Phi|}{\alpha})-\frac{m^2\,\eta^2_{L}}{\tilde{\omega}^2}.
\label{31}
\end{eqnarray}
Equation (\ref{31}) is the eigenvalue of a charged scalar particle in the presence of external fields including  an internal magnetic flux field in a cosmic string space-time with a spacelike dislocation subject to a linear scalar potential. For $\alpha \rightarrow 1$, the energy eigenvalue Eq. (\ref{31}) reduces to the result obtained in Ref. \cite{aa30} (see Eq. (42) in Ref. \cite{aa30}). With $\chi=0$, we have that the energy eigenvalue (\ref{31}) becomes that stem from the interaction of the Klein-Gordon oscillator with a magnetic field and a linear scalar potential in the Minkowski space-time with a cosmic string. Thus, the topology of the space-time also changes the pattern of oscillation of the ground state energies.

\vspace{0.5cm}
{\bf Case 2 :} Absence of external magnetic field, $B_0=0$.
\vspace{0.5cm}

We choose here $B_0 \rightarrow 0$, that is, no external fields into the considered relativistic quantum system. In that case, the energy eigenvalue Eq. (\ref{29}) becomes
\begin{eqnarray}
&&E^2_{n,l}=m^2+k^2+2\,\sqrt{m^2\,\Omega^2+\eta^2_{L}}\,(n+1+\sqrt{\frac{(l-k\,\chi-\Phi)^2}{\alpha^2}+\eta^2_{c}})\nonumber\\
&&-\frac{m^2\,\eta^2_{L}}{m^2\,\Omega^2+\eta^2_{L}}+2\,\eta_c\,\eta_L+2\,m\,\Omega.
\label{32}
\end{eqnarray}
Equation (\ref{32}) is the energy eigenvalue of Klein-Gordon oscillator field in the presence of an internal magnetic flux field is a cosmic string space-time with a spacelike dislocation subject to a Cornell-type scalar potential. For $\alpha \rightarrow 1$, the energy eigenvalue Eq. (\ref{32}) reduces to the result obtained in Ref. \cite{aa31}. With $\chi=0$, we have that the energy eigenvalue (\ref{32}) becomes that stem from the interaction of the Klein-Gordon oscillator with a Cornell-type scalar potential in the Minkowski space-time with a cosmic string. Thus, the topology of the space-time also changes the pattern of oscillations of the ground state energies.

\vspace{0.5cm}
{\bf Case 3 :} Zero dislocation parameter $\chi=0$.
\vspace{0.5cm}

We choose here zero torsion parameter, $\chi \rightarrow 0$ into the considered relativistic system. In that case, the energy eigenvalue Eq. (\ref{29}) becomes
\begin{eqnarray}
&&E^2_{n,l}=m^2+k^2+2\,\eta_c\,\eta_L+2\,m\,\Omega+\frac{2\,m\,\omega_c}{\alpha}\,(l-\Phi)\nonumber\\
&&+2\,\tilde{\omega}\,(n+1+\sqrt{\frac{(l-\Phi)^2}{\alpha^2}+\eta^2_{c}})-\frac{m^2\,\eta^2_{L}}{\tilde{\omega}^2}.
\label{36}
\end{eqnarray}
Equation (\ref{36}) is the energy eigenvalue of a massive charged particle in the presence of external uniform magnetic field including an internal magnetic flux subject to a Cornell-type scalar potential in a cosmic string space-time. 

In all the above cases, we see that the relativistic energy eigenvalue depends on the geometric quantum phase \cite{cc1} which gives rise to the analogous effect to the Aharonov-Bohm effect for a bound states. Besides, we have that, $E_{n,{\bar l}} (\Phi_B+\Phi_0)=E_{n,{\bar l} \mp \tau} (\Phi_B)$ where, $\Phi_0=\pm\,\frac{2\,\pi\,\alpha}{e}\,\tau$ with $\tau=1,2,3,.......$ and ${\bar l}=\frac{l}{\alpha}$. It is observed in {\it case 1}-{\it case 2} that the angular momentum eigenvalue $l$ is shifted, $l\rightarrow l_{eff}=\frac{1}{\alpha}\,(l-\Phi-k\,\chi)$ whereas in {\it case 3}, it is $l\rightarrow l'=\frac{1}{\alpha}\,(l-\Phi)$, an effective angular quantum number. As done earlier one can evaluate the individual energy level and eigenfunction one by one.

\section{Conclusions}

We have investigated the effect of torsion and topological defects that stem from a space-time with a spacelike dislocation on the interactions between an electrically charged particle and an external uniform magnetic field. Besides, we have assumed that the topological defects have an internal magnetic flux flux. By solving the Klein-Gordon oscillator equation analytically in {\it sub-section 2.1}, we have obtained the relativistic energy eigenvalue Eq. (\ref{b6}) and corresponding eigenfunction Eq. (\ref{b7}). We have shown that for $\alpha \rightarrow 1$, the energy eigenvalue reduces to the result obtained in Ref. \cite{aa30}. We have seen in Eq. (\ref{b6}) that there exists an effective angular momentum quantum number, $l \rightarrow l_{eff}=\frac{1}{\alpha}\,(l-k\,\chi-\frac{\phi_B}{(2\,\pi/e)})$. Thus the relativistic energy eigenvalue depend on the geometric quantum phase \cite{cc1}. This dependence of the energy eigenvalue on the geometric quantum phase gives rise to the analogue effect to the Aharonov-Bohm effect for bound states \cite{cc2,aa48,aa49}. Thus, we have that $E_{n,{\bar l}} (\Phi_B+\Phi_0)=E_{n,{\bar l} \mp \tau} (\Phi_B)$ where, $\Phi_0=\pm\,\frac{2\,\pi\,\alpha}{e}\,\tau$ with $\tau=1,2,3,.......$ and ${\bar l}=\frac{l}{\alpha}$. For zero torsion parameter ($\chi \rightarrow 0$), we have also obtained the energy eigenvalue Eq. (\ref{b9}) which is the extended result in comparison to those obtained in \cite{aa7} in a cosmic string space-time in the presence of external fields including an internal magnetic flux field. We have seen that the presence of torsion ($\chi \neq 0$) in the space-time modifies the degeneracy of the relativistic energy levels. Besides, the presence of torsion in the space-time changes the pattern of oscillation of the energy levels. 

We have extended our above discussion to investigate the behaviour of this relativistic system under the influence of a Cornell-type scalar potential in {\it sub-section 2.2}. We have solved the Klein-Gordon oscillator equation in the cosmic string space-time with a spacelike dislocation and obtained the energy eigenvalue Eq. (\ref{29}). We have seen that for $\alpha \rightarrow 1$ and $\eta_c \rightarrow 0$, this energy eigenvalue reduces to the result obtained in Ref. \cite{aa30}. Furthermore, in the absence of external fields ($B_0 \rightarrow 0$), this energy eigenvalue Eq. (\ref{29}) reduces to the result obtained in Ref. \cite{aa31}. Thus, we have observed that the relativistic energy eigenvalue Eq. (\ref{29}) is the extended results in comparison to those obtained in Refs. \cite{aa30,aa31}. Also, we have seen that the relativistic energy eigenvalue Eq. (\ref{29}) depends on the Aharonov-Bohm geometric quantum phase \cite{cc1}. This dependence of the relativistic energy eigenvalue on the geometric quantum phase gives rise to the analogue effect to the Aharonov-Bohm effect for bound states \cite{cc2,aa48,aa49}. We have that $E_{n,{\bar l}} (\Phi_B+\Phi_0)=E_{n,{\bar l} \mp \tau} (\Phi_B)$ where, $\Phi_0=\pm\,\frac{2\,\pi\,\alpha}{e}\,\tau$ with $\tau=1,2,3,.......$ and ${\bar l}=\frac{l}{\alpha}$. Thus we have seen that the presence of torsion ($\chi \neq 0$) in the space-time modifies the degeneracy of the relativistic energy levels. Besides, the presence of torsion in the space-time changes the pattern of oscillation of the energy levels.  For $\chi=0$, we have also obtained the relativistic energy eigenvalue Eq. (\ref{36}) of a massive charged particle in the presence of external fields including an internal magnetic flux field in a cosmic string space-time subject to a Cornell-type scalar potential. We have seen that the energy eigenvalue depends on the geometric quantum phase \cite{cc1} which gives rise to the analogue effect to the Aharonov-Bohm effect for bound states \cite{cc2,aa48,aa49}. 

In recent years, thermodynamic properties of quantum systems \cite{bb18,bb19,bb20,bb21,bb22}, quantum Hall effect \cite{aa23,bb23,bb24}, displaced Fock states \cite{bb25,bb26}, and the possibility of building coherent state \cite{bb27,bb28,bb29,bb30} have attracted a great current research interest in the literature. It is well known in non-relativistic quantum mechanics that the Landau quantization is the simplest system that we can work with the studies of quantum Hall effect. Therefore, the relativistic systems analyzed in this work may be used for investigating the influence of torsion, topological defects (cosmic string) as well as the potential for searching the relativistic analogue to the quantum Hall effect, coherent states, and displaced Fock states in topological defects space-time with a spacelike dislocation. So the results are given in this paper with those in Refs. \cite{aa7,aa30,aa31} would present the above interesting effects.

\section*{Acknowledgement}

Author sincerely acknowledged the anonymous kind referee(s) for their valuable comments and suggestions.

\section*{Data Availability}

There is no data associated with this manuscript or no data has been used to prepare this manuscript.

\section*{Conflict of Interest}

Author declares that there is no conflict of interest regarding publication this paper.

\end{document}